\documentclass{article}

\usepackage[english]{babel}

\usepackage[letterpaper,top=2cm,bottom=2cm,left=3cm,right=3cm,marginparwidth=1.75cm]{geometry}

\usepackage[colorlinks=true, allcolors=blue]{hyperref}
\usepackage{amsmath}
\usepackage{amsthm,amsmath,amssymb}
\usepackage[sort&compress]{natbib}
\setcitestyle{numbers,square}

\usepackage{mathrsfs}
\usepackage{algorithm}
\usepackage{algorithmic}
\usepackage{appendix}
\usepackage{graphicx}
\usepackage{subfigure}
\usepackage{diagbox}
\usepackage{hhline}
\usepackage{multirow}
\usepackage{bm}
\usepackage{doi}

\newtheorem{lemma}{Lemma}
\newtheorem{pof}{Proof}
\newtheorem{theorem}{Theorem}

\newtheorem{remark}{Remark}

\newcommand{\uu}{\mathbf{u}}
\newcommand{\xx}{\mathbf{x}}
\newcommand{\PP}{\mathcal{P}}
\newcommand{\FF}{\mathcal{F}}
\newcommand{\bb}{\mathbf{b}}
\newcommand{\RR}{\mathbb{R}}
\newcommand{\vv}{\mathbf{\tilde{u}}}
\newcommand{\ww}{\mathbf{w}}
\newcommand{\GG}{\mathbf{G}}
\newcommand{\zz}{\mathbf{z}}
\newcommand{\ZZ}{\mathcal{Z}}
\newcommand{\lamb}{\bm{\lambda}}

\title{ADMM based Fourier phase retrieval with untrained generative prior}
\author{Liyuan Ma \thanks{
College of Science, National University of Defense Technology,
Changsha, Hunan, 410073, P.R.China. Email: \texttt{maliyuan21@nudt.edu.cn}}
\and Hongxia Wang {\thanks{
College of Science, National University of Defense Technology,
Changsha, Hunan, 410073, P.R.China. Corresponding author. Email: \texttt{wanghongxia@nudt.edu.cn}}}
\and Ningyi Leng {\thanks{
College of Science, National University of Defense Technology,
Changsha, Hunan, 410073, P.R.China. Email: \texttt{lengningyi14@nudt.edu.cn}}}
\and Ziyang Yuan {\thanks{Academy of Military Science of People’s Liberation Army, Beijing, P.R.China. Email: \texttt{yuanziyang11@nudt.edu.cn}}}
}

\date{}

\begin{document}
\maketitle

\begin{abstract}
Fourier phase retrieval (FPR) is an inverse problem that recovers the signal from its Fourier magnitude measurement, it's ill-posed especially when the sampling rates are low. In this paper, an untrained generative prior is introduced to attack the ill-posedness. Based on the alternating direction method of multipliers (ADMM), an algorithm utilizing the untrained generative network called Net-ADM is proposed to solve the FPR problem. Firstly, the objective function is smoothed and the dimension of the variable is raised to facilitate calculation. Then an untrained generative network is embedded in the iterative process of ADMM to project an estimated signal into the generative space, and the projected signal is applied to next iteration of ADMM. We theoretically analyzed the two projections included in the algorithm, one makes the objective function descent, and the other gets the estimation closer to the optimal solution. Numerical experiments show that the reconstruction performance and robustness of the proposed algorithm are superior to prior works, especially when the sampling rates are low.

$\mathbf{Keywords}$:~~Fourier phase retrieval,~~untrained generative prior,~~alternating direction method of multipliers
\end{abstract}

\section{Introduction}
Fourier phase retrieval (FPR) seeks to recover an unknown signal $\xx \in \RR^n$ from its Fourier magnitude measurement $\bb = |\FF \xx| + \eta \in \RR^m \mbox{ }(m>n)$, where $\eta$ denotes additive noise. It arises in various applications, such as X-ray crystallography \cite{Jianwei2008,Pinilla2018}, ptychography \cite{Chen2017}, diffraction imaging \cite{Luke2005} and astronomical imaging \cite{Dainty1987}. It is in general ill-posed since there are many different signals which have the same Fourier magnitude \cite{Fannjiang2020}. For $\xx \in \RR^n$, a theoretical work \cite{Balan2006} indicates that the uniqueness of the solution can be guaranteed when $m \ge 2n-1$ and every subset of $n$ measurement vectors
spans $\RR^n$. Here we use sampling rate $r=m/n$ to describe the ratio of measurement length to signal length in each dimension. The signal is difficult to be reconstructed when the sampling rate is low generally.

The models of addressing phase retrieval (PR) problems can be divided into convex models and non-convex models. Convex models relax the non-convex problem into convex one, such as PhaseLift \cite{Cand2013}, PhaseMax \cite{Goldstein2018}, etc. We consider the non-convex quadratic model $\underset{\xx}{min}\mbox{ }\frac{1}{2m}\Vert \bb - |\FF \xx| \Vert^2$ because that extensive numerical and experimental validation confirms that the magnitude-based cost function performs significantly better than the intensity-based one \cite{Yeh2015, Wang2018}. Based on the model, it's common to add some prior information about the signal $\xx$ to relieve the ill-posedness, i.e.
\begin{equation} \label{equ1}
    \underset{\xx \in \mathbb{S}}{\mathop{\min}}\quad \frac{1}{2m}\Vert \bb - |\FF \xx| \Vert^2,
\end{equation}
where $\mathbb{S}$ is a set possibly constrained by priors. The common prior information is support constraint \cite{Chen2007}, which considers that the signal outside boundaries is zero. Sparse prior is also used for FPR to recover the desired signal using a minimal number of measurements \cite{Sarangi2017}, the signal has few non-zero coefficients in some basis under this assumption.

Recently, deep generative priors \cite{Hand2018,Hyder2019,Shamshad2018,Jagatap2019,RakibHyder2019} which assume the signal is within the range of a generative network are proposed to replace hand-crafted priors. Among them, trained generative networks \cite{Hand2018,Hyder2019,Shamshad2018} build priors by learning the distribution information of signal from massive amounts of training data, which achieve superior results. However, in many scenarios we can't obtain sufficient training data, and there may even be cases where the reconstruction performance of network is poor due to inconsistent data distribution, so it's necessary to introduce untrained generative priors. The untrained generative networks such as Deep Image Prior (DIP) \cite{Ulyanov2017} and its variant Deep Decoder \cite{Heckel2019} are proposed, they don't rely on any external data to learn the distribution information of the signal, because 
their structure can capture the low-level statistical priors of images implicitly \cite{Ulyanov2017}.

Although a variety of excellent PR models have been proposed, designing efficient algorithms to solve these models is still challenging. Algorithms to solve PR problem can be divided into classical algorithms and learning-based algorithms. The earliest classical methods are based on alternating projections, such as Gerchberg-Saxton (GS) \cite{Gerchberg1972} and its variant, hybrid-input-output (HIO) \cite{Fienup1982}. The algorithms based on gradient descent are proposed later, such as Wirtinger Flow (WF) \cite{Candes2014}, Truncated Wirtinger Flow (TWF) \cite{Chen2015} and related variants \cite{Yuan2019}. In addition, there are also algorithms based on second-order derivatives \cite{Ma2018}, which are faster than WF. Most of the classical algorithms are based on the above algorithms or their variants.

With the development of neural networks, learning-based algorithms gain a lot of momentum. There are some algorithms which take deep neural networks as denoisers in the iterative process of PR \cite{Metzler2018, Isil2019}, and other algorithms utilize deep neural networks to learn a mapping that reconstructs the images from measurements \cite{Sinha2017, Cherukara2018}. Deep generative networks as image priors have been recently introduced into PR problem. There are some algorithms \cite{Hand2018,Hyder2019,Shamshad2018} that use trained or pre-trained generative networks to solve the Fourier or Gaussian PR problems (which replaces Fourier matrix $\FF$ with a Gaussian random matrix). However, it's inevitable that their reconstruction performance depends on massive amounts of training data.

The algorithms \cite{Jagatap2019, jagatapGauri2019} based on untrained generative priors named Net-GD and Net-PGD have been proposed recently and applied to Gaussian PR, they adopt gradient descent and projected gradient descent respectively. Even though there have been many researches about Gaussian PR, FPR is more challenging \cite{bandeira2014} and applicable in a wide range of scenarios. When Net-GD and Net-PGD are applied to FPR problem, their reconstruction performance and robustness are not ideal, even when the sampling rates are high. In summary, the application of untrained generative networks to FPR problems remains to be further explored, including theoretical guarantees and algorithms that can be adopted at low sampling rates.

This paper proposes an algorithm named Net-ADM that combines the alternating direction method of multipliers (ADMM) with untrained generative network to tackle FPR problem. Specifically, we smooth the amplitude-based loss function (\ref{equ1}) and improve the dimension of the variable, so that the objective function is Lipschitz differential and the Fourier transform can be calculated by fast Fourier transform (FFT) \cite{Potts2008}. Then the FPR problem is modeled as an constrained optimization problem and ADMM is adopted to solve the optimization problem alternately. Additionally, the untrained generative network is embedded into the iterative process to project an estimated signal of the ADMM into the generative space that captures some image characteristics, the projected signal can be applied to next iteration of ADMM. We theoretically analyze the two projections included in Net-ADM algorithm, one makes the objective function descent, and the other makes the estimation closer to the optimal solution, these good properties may be the reason for the gain compared to ADMM. Numerical experiments show that the reconstruction performance and robustness of Net-ADM are superior as compared to prior works, especially when the sampling rates are low.

The remainder of the paper is organized as follows. In section \ref{Net-ADM}, we establish the FPR model based on untrained generative prior, describe the Net-ADM algorithm and give relevant theoretical analysis. Section \ref{experiment} describes the experimental settings, conducts numerical experiments under different sampling rates and noise levels, the effect of parameters in the algorithm is discussed finally. In section \ref{conclusion}, some conclusions are drawn and new ideas are proposed for future work. 

In this paper, the bold lowercase letters or numbers represent vectors, calligraphic uppercase letters denote operators and the uppercase letters on the blackboard bold represent spaces or sets. $\sqrt{\cdot}$, $|\cdot|$ and $\div$ are element-wised operators. $\Vert \cdot \Vert$ is the Euclidean norm of a vector or the spectral norm of a matrix. Hadamard product is represented by $\odot$. Vectorization of a matrix is written as $vec(\cdot)$. $(\cdot)_{[n+1:m]}$ denotes a sub-vector constructed by the last $m-n$ elements.

\section{Net-ADM} \label{Net-ADM}

\subsection{FPR Model based on untrained generative prior}

The untrained generative network incorporates implicit priors about the signals they generate, and these prior assumptions are built into the network structure. Thus we expand the components of the network in detail and analyze the prior information that may be carried by its structure.

Under the untrained generative prior, $\mathbb{S}$ in (\ref{equ1}) represents the range of a network which is denoted by $\GG(\ww;\zz)$. The input $\zz\in \RR^{d} (d \ll n)$ of network represents the low-dimensional latent code, its elements are fixed and generated from uniform random distribution; $\ww$ represents the weights that need to be learned but not pre-trained. 
This paper draws on the network structure in \cite{jagatapGauri2019}, which exchanges the order of upsampling with a composition of activation function and channel normalization compared to deep decoder \cite{Heckel2019}. 

For a $J$-layer network, we denote the input of $j+1$-th layer as $Z_{j} \in \RR^{c_{j} \times d_{j}}$ ($j = 0, 1, \cdots, J-1$) and $Z_J\in \RR^{c_J\times d_J}$ represents the output of $J$-th layer, in addition, $\zz = vec(Z_0)$ and $d = c_0 \times d_0$. The network consists of $1\times 1$ convolutions' weights $W_j \in \RR^{c_{j+1} \times c_{j}}$, ReLU activation function $relu(\cdot)$, channel normalization operator $cn(\cdot)$ and bi-linear upsampling operators $U_{j}$. The operators from the last layer to the output includes the $1\times 1$ convolutions' weight $W_J \in \RR^{c_{out}\times c_J}$ and sigmoid activation function $sig(\cdot)$, where $c_{out} = 1$ for a grayscale image and $c_{out} = 3$ for an RGB image. Thus, the network at layer $j+1$ can be expressed as
\begin{equation} \label{net}
    Z_{j+1} = U_j cn\left( relu\left(W_{j} Z_{j}\right)\right), \quad j = 0, 1, \cdots, J-1.
\end{equation}
The output of network is $G (\ww;\zz) = sig\left(W_J Z_J\right)$, where $\ww = \{W_0, W_1, \cdots, W_J\}$ are weights to learn. It's apparent that the manually adjustable parameters of the network are network depth and number of weight channels in each layer. We can use $\{c_0, c_1, \cdots, c_J\}$ to represent the designed structure for a $J$-layer network. Additionally, simplifying (\ref{net}) gives
\begin{equation} \label{sim-net}
    G(\ww;\zz) = \left(\sigma_J\circ \ZZ_{J} \circ \sigma_{J-1} \circ \ZZ_{J-1} \circ \cdots \circ \sigma_0 \circ \ZZ_0\right) (\ww),
\end{equation}
where $\ZZ_j\mbox{ }(j = 0, 1, \cdots, J)$ denote the linear operators in the network because the $1 \times 1$ convolution which is the operation between $\ZZ_j$ and $\ww$ can be viewed as a linear combination of their channels, $\sigma_j = U_j \circ cn \circ relu \mbox{ } (j = 0, 1, \cdots, J-1)$ denote the composite operators of ReLU activation function, channel normalization operator and bi-linear upsampling, $\sigma_J = sig$ denotes the sigmoid activation function. These operators are fixed, thus the network can be regarded as a function of $\ww$.

\begin{remark}
    Since the input $Z_0 \in \RR^{c_0 \times d_0}$ is randomly generated from the [0, 0.1] uniform distribution, it spans $\RR^{c_0 \times d_0}$ with probability 1. The structure of the network restricts the representation space of $G(\ww;\zz)$ to a certain range $\mathbb{S}$, which captures some characteristics of images. For example, the bi-linear upsampling operators $U_j\mbox{ } (j = 0, 1, \cdots, J-1)$ reflect the correlation of adjacent pixels in an image. Specifically, which image in the space $\mathbb{S}$ is represented by the network is determined by $\ww$.
\end{remark}

Our goal is to find optimal weights $\ww^{\ast}$ to represent the optimal solution $\xx^{\ast}$, which is equivalent to substituting the surjective mapping \cite{jagatapGauri2019} $G:\ww \to \xx$ and minimizing the loss function (\ref{equ2}) over $\ww$,
\begin{equation} \label{equ2}
    \underset{\ww \in \mathbb{W}}{\mathop{\min}}\mbox{ } \psi(G(\ww;\zz))\mbox{, } \mbox{where } \psi (\xx) = \frac{1}{2m}\Vert \bb - |\FF \xx| \Vert^2.
\end{equation}
where $\mathbb{W}$ denotes the set of weights. Inspired by \cite{Li2020}, we can obtain the existence of the solution of model (\ref{equ2}). 

\begin{theorem}
    For the network (\ref{sim-net}), if the set of weights $\mathbb{W}$ is a bounded weakly closed subset in reflexive Banach space, there at least exists a solution $\ww^{\ast}$ of model (\ref{equ2}).
\end{theorem}
\begin{pof}
    Since the form of (\ref{equ2}) and (\ref{sim-net}) is similar to (1.2) and (2.1) of \cite{Li2020} respectively, they satisfy the conditions similar to (A1) of Condition 2.2, which guarantees the weakly lower semicontinuity of the objective function $F(\ww) = \psi(G(\ww;\zz))$. Specifically, since the elements in $\zz$ belong to $[0, 0.1]$ generally, $\mathcal{Z}_j$ are bounded linear when $\mathbb{W}$ is bounded. 
    In addition, $\sigma_j$ are weakly continuous, and the function $\psi(\cdot) = \frac{1}{2m} \Vert \bb - |\FF(\cdot)| \Vert^2$ is weakly lower semi-continuous. Thus, the objective function $F(\ww)$ is weakly lower semi-continuous. For $\ww \in \mathbb{W}$, when $\mathbb{W}$ is a bounded weakly closed subset in reflexive Banach space, $F(\ww)$ is able to reach infimum on $\mathbb{W}$, thus there at least exits a solution $\ww^{\ast}$ of model (\ref{equ2}).
\end{pof}

Note that the conclusion in \cite{Li2020} assumes all free parameters in $F(\ww)$ are trained before minimization of (\ref{equ2}), it is also satisfied here since the untrained generative network (\ref{net}) does not require to be pre-trained and all operators are fixed. Furthermore, theorem 1 requires that the set of weights $\mathbb{W}$ is a closed set. We can use strategies such as weight decay and so on in the learning process of network to ensure that the condition is satisfied.

For the convenience of algorithm implementation, we rewrite the model (\ref{equ2}) as the following optimization model by using the intermediate variable $\xx$.
\begin{equation} \label{equ13}
    \begin{split}
        \underset{\xx, \ww}{\mathop{\min}} & \mbox{ } \psi (\xx), \\
        s.t. & \mbox{ } \xx - G(\ww;\zz) = 0.
    \end{split}
\end{equation}

\subsection{Algorithm design}

In this section, we will design an algorithm to solve (\ref{equ13}), which combines the ADMM with untrained generative network, thus it's called Net-ADM method in this paper. Firstly, in order to execute Fourier transform by FFT algorithm, we introduce an auxiliary variable $\uu \in \RR^m\mbox{ } (m > n)$ whose last $m-n$ elements are zeros $\uu_{[n+1:m]}=0$ such that $\FF$ represents FFT operator. Secondly, in order to make the objective function $\psi(\cdot)$ Lipschitz differential, inspired by Chang \cite{Chang2019}, we add a penalty $\varepsilon \mathbf{1}$ on both sides of $\bb = |\FF\uu|$, then use radical smoothing, the loss function $\psi(\xx)$ can be rewritten as follows:
\begin{equation}
    f(\uu) = \frac{1}{2m} \Vert \sqrt{\bb^2 + \varepsilon \mathbf{1}} - \sqrt{\left| \FF\uu \right|^2 + \varepsilon \mathbf{1}} \Vert^2,
\end{equation}
where $\varepsilon >0$ represents penalization parameter, $\mathbf{1} \in \RR^m$ represents a vector whose elements are all ones.
The gradient of $f(\uu)$ is shown in (\ref{equ5}), its continuity is proved in Lemma \ref{lemma1}.
\begin{equation} \label{equ5}
    \nabla f(\uu) = \uu - \FF^{-1}\left( \frac{\sqrt{\bb^2+\varepsilon \mathbf{1}}}{\sqrt{|\FF(\uu)|^2+\varepsilon \mathbf{1}}} \odot \FF(\uu) \right).
\end{equation}
\begin{lemma} \label{lemma1}
    The function $f(\uu)$ is gradient Lipschitz continuous,
    \begin{equation} \label{equ8}
        \Vert \nabla f(\uu_2) - \nabla f(\uu_1) \Vert \le L \Vert \uu_2 - \uu_1 \Vert,\mbox{ } \forall \uu_1, \uu_2 \in \RR^m
    \end{equation}
    where $L = 1 + \frac{2}{\sqrt{\varepsilon}} \Vert \sqrt{\bb^2+\varepsilon \mathbf{1}} \Vert_\infty$.
\end{lemma}

\begin{pof}
    \begin{equation}
        \begin{split}
            \quad & \Vert \nabla f(\uu_2) - \nabla f(\uu_1) \Vert\\
            \quad & = \Vert \uu_2 - \FF^{-1}\left( \frac{\sqrt{\bb^2+\varepsilon \mathbf{1}}}{\sqrt{|\FF \uu_2|^2 + \varepsilon \mathbf{1}}} \odot \FF \uu_2 \right) -\uu_1 + \FF^{-1}\left( \frac{\sqrt{\bb^2+\varepsilon \mathbf{1}}}{\sqrt{|\FF \uu_1|^2 + \varepsilon \mathbf{1}}} \odot \FF \uu_1 \right) \Vert\\
            \quad & \le \Vert \uu_2 - \uu_1 \Vert + \Vert \FF^{-1}\left( \frac{\sqrt{\bb^2+\varepsilon \mathbf{1}}}{\sqrt{|\FF \uu_2|^2 + \varepsilon \mathbf{1}}} \odot \FF \uu_2 - \frac{\sqrt{\bb^2+\varepsilon \mathbf{1}}}{\sqrt{|\FF \uu_1|^2 + \varepsilon \mathbf{1}}} \odot \FF \uu_1 \right) \Vert \\
            \quad & = \Vert \uu_2 - \uu_1 \Vert + \frac{1}{\sqrt{m}} \Vert \frac{\sqrt{\bb^2+\varepsilon \mathbf{1}}}{\sqrt{|\FF \uu_2|^2 + \varepsilon \mathbf{1}}} \odot \FF \uu_2 - \frac{\sqrt{\bb^2+\varepsilon \mathbf{1}}}{\sqrt{|\FF \uu_1|^2 + \varepsilon \mathbf{1}}} \odot \FF \uu_1 \Vert \mbox{    {\rm (by Parseval's theorem)}} \\
            \quad & \le \Vert \uu_2 - \uu_1 \Vert + \frac{1}{\sqrt{m}} \Vert \frac{\sqrt{\bb^2+\varepsilon \mathbf{1}}}{\sqrt{|\FF \uu_2|^2 + \varepsilon \mathbf{1}}} \odot \left( \FF \uu_2 -  \FF \uu_1 \right) \Vert \\
            \quad & + \frac{1}{\sqrt{m}} \Vert \left( \frac{\sqrt{\bb^2+\varepsilon \mathbf{1}}}{\sqrt{|\FF \uu_2|^2 + \varepsilon \mathbf{1}}} - \frac{\sqrt{\bb^2+\varepsilon \mathbf{1}}}{\sqrt{|\FF \uu_1|^2 + \varepsilon \mathbf{1}}} \right) \odot \FF \uu_1 \Vert \\
            \quad & \le \Vert \uu_2 - \uu_1 \Vert + \frac{2}{\sqrt{\varepsilon}} \Vert \sqrt{\bb^2+\varepsilon \mathbf{1}} \Vert_\infty \Vert \uu_2 - \uu_1 \Vert
        \end{split}
    \end{equation}
    Among them, 
    \begin{equation}
        \begin{split}
            \quad & \frac{1}{\sqrt{m}} \Vert \left( \frac{\sqrt{\bb^2+\varepsilon \mathbf{1}}}{\sqrt{|\FF \uu_2|^2 + \varepsilon \mathbf{1}}} - \frac{\sqrt{\bb^2+\varepsilon \mathbf{1}}}{\sqrt{|\FF \uu_1|^2 + \varepsilon \mathbf{1}}}\right) \odot \FF \uu_1 \Vert\\
            \quad & = \frac{1}{\sqrt{m}} \Vert \sqrt{\bb^2+\varepsilon \mathbf{1}} \odot \frac{\FF \uu_1 \odot \left(|\FF \uu_1|^2 - |\FF \uu_2|^2 \right)}{\sqrt{|\FF \uu_2|^2 + \varepsilon \mathbf{1}} \sqrt{|\FF \uu_1|^2 + \varepsilon \mathbf{1}} \left(\sqrt{|\FF \uu_1|^2 + \varepsilon \mathbf{1}} + \sqrt{|\FF \uu_2|^2 + \varepsilon \mathbf{1}}\right)} \Vert \\
            & \le \frac{1}{\sqrt{\varepsilon}}\Vert \sqrt{\bb^2+\varepsilon \mathbf{1}} \Vert_\infty \Vert \uu_2 - \uu_1 \Vert,
        \end{split}
    \end{equation}
    Therefore, $\Vert \nabla f(\uu_2) - \nabla f(\uu_1) \Vert \le L \Vert \uu_2 - \uu_1 \Vert$, where $L = 1+\frac{2}{\sqrt{\varepsilon}} \Vert\sqrt{\bb^2 + \varepsilon \mathbf{1}} \Vert_\infty$.
\end{pof}
Note that $\uu_{[n+1:m]}=0$ is equivalent to $\uu - \PP \PP^T \uu = 0$. The optimization model (\ref{equ13}) now turns to
\begin{subequations} \label{equ3}
    \begin{align}
        \underset{\uu, \ww}{\mathop{\min}} &\mbox{ } f(\uu), \label{min} \\
        s.t. & \mbox{ } \PP \PP^T \uu - \uu = 0 \label{con1}, \\
        \quad & \mbox{ } \PP^T\uu - \GG(\ww; \zz) = 0. \label{con2}
    \end{align}
\end{subequations}
where $\PP:\RR^n \to \RR^m$ denotes zero padding operator, i.e. it appends $m-n$ zero elements after the last element of the vector. And $\PP^T:\RR^m \to \RR^n$ means to obtain the first $n$ elements of the vector. 
The solution of (\ref{equ13}) is $\xx^{\ast} = \PP^T \uu^{\ast}$, if $\uu^{\ast}$ is a solution of (\ref{equ3}).

\begin{figure}[htbp]
    \centering
    \includegraphics[height = 5.2 cm]{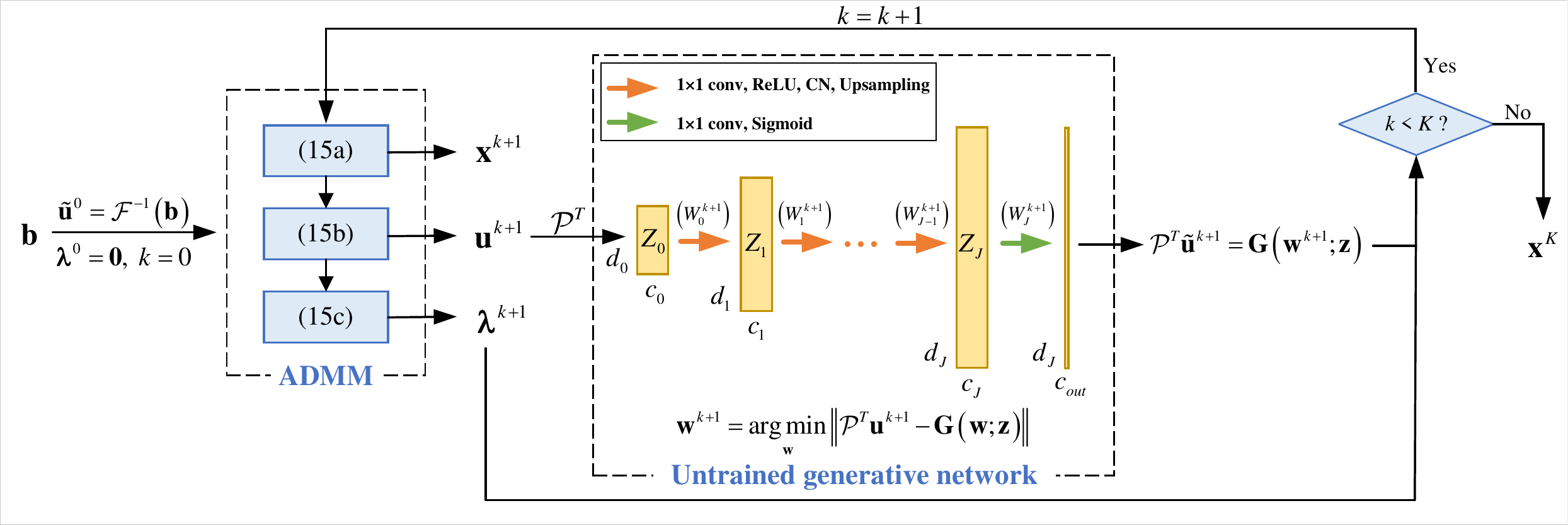}
    \caption{Illustration of Net-ADM algorithm. It consists of two main parts: ADMM and an untrained generative network. ADMM is used to solve the sub-problem (\ref{sub1}), the specific schemes are shown in (\ref{ADMM:x})-(\ref{ADMM:lambda}). Then an untrained network is adopted to solve the sub-problem (\ref{sub2}), the estimated signal $\PP^T\hat{\uu}^{k+1}$ of ADMM is projected to the range $\mathbb{S}$ of the network to obtain $\PP^T\vv^{k+1}$, as shown in (\ref{ADMM:weight}) and (\ref{ADMM:net}). Where $\mathbb{S}$ captures some characteristics of images and $\PP^T\vv$ can be applied in next iteration of ADMM. The network is particularly simple, which is roughly similar to deep decoder, as each layer has the same structure, consisting of $1\times 1$ convolutions, channel-wise normalization (CN), ReLU activation functions and upsampling operators.}
    \label{fig: net-admm}
\end{figure}

We divide (\ref{equ3}) into two sub-problems and solve them alternately. Firstly, taking the objective function (\ref{min}) and the first constraint (\ref{con1}) into consideration, we replace the variable $\PP^T\uu$ with $\xx \in \RR^n$ in (\ref{con1}), then ADMM can be adopted to solve the following sub-problem alternately.
\begin{equation} \label{sub1}
    \left(\hat{\uu}, \hat{\xx}\right) = \underset{\uu, \xx}{\mathop{\arg\min}} \mbox{ } f(\uu), \mbox{ } s.t. \mbox{ } \PP \xx - \uu = 0.
\end{equation}
Secondly, we transform the second constraint (\ref{con2}) into the second sub-problem. The untrained generative network approximates the intermediate variable $\PP^T\hat{\uu}$ of ADMM, aiming to project it into the range of $\GG(\ww;\zz)$, which captures some characteristics of images.
\begin{equation} \label{sub2}
    \PP^T \tilde{\uu} = \GG(\hat{\ww};\zz), \mbox{ where } \hat{\ww} = \underset{\ww}{\mathop{\arg\min}} \mbox{ } \Vert \PP^T\hat{\uu} - \GG(\ww; \zz) \Vert.
\end{equation}
We can get weights $\hat{\ww}$ to represent the substitute variable $\PP^T \vv = \GG(\hat{\ww};\zz)$ of $\PP^T\hat{\uu}$, then $\PP^T \vv$ is applied to next iteration of ADMM. The specific execution process can be found in the illustration of the algorithm, as shown in Fig.\ref{fig: net-admm}.

According to augmented Lagrange multiplier method, we can transform (\ref{sub1}) into an unconstrained optimization problem:
\begin{equation}\label{equ6}
    L_\rho (\xx, \uu, \lamb) = f(\uu) + \langle \lamb, \PP \xx - \uu \rangle + \frac{\rho}{2} \Vert \PP \xx - \uu \Vert^2,
\end{equation}
where $\lamb \in \RR^m$ is the Lagrangian multiplier associated with the equality constraint, $\rho > 0$ is the coefficient of the quadratic penalty term. Therefore, we can minimize $L_\rho (\xx, \uu, \lamb)$ with respect to $\uu$ and $\xx$ in an alternating way, which is the basic idea of ADMM. After obtaining the intermediate variable $\hat{\uu}$ of the ADMM, the untrained generative network is used to optimize (\ref{sub2}) over $\ww$. The variable $\PP^T \vv = \GG(\hat{\ww};\zz)$ represented by the network is applied to next iteration. Specifically, given some initial vectors $\vv^0, \lamb^0$, we can update the variables in the following iterative format:
\begin{subequations} \label{net-adm}
    \begin{align}
        \xx^{k+1}& = \underset{\xx}{\mathop{\arg\min}}\left\{ \langle \lamb^k, \PP \xx - \vv^{k} \rangle + \frac{\rho}{2} \Vert \PP \xx - \vv^{k} \Vert^2 \right\}, \label{ADMM:x} \\
        \uu^{k+1}& = \underset{\uu}{\mathop{\arg\min}} \left\{f(\uu) + \langle \lamb^k, \PP \xx^{k+1} - \uu \rangle + \frac{\rho}{2} \Vert \PP \xx^{k+1} - \uu \Vert^2 \right\}, \label{ADMM:u} \\
        \lamb^{k+1}& = \lamb^k + \rho \left(\PP\xx^{k+1} - \uu^{k+1}\right), \label{ADMM:lambda} \\
        \ww^{k+1} & = \underset{\ww}{\mathop{\arg\min}} \Vert \PP^T\uu^{k+1} - \GG(\ww;\zz) \Vert, \label{ADMM:weight} \\
        \PP^T\vv^{k+1} & = \GG(\ww^{k+1};\zz). \label{ADMM:net}
    \end{align}
\end{subequations}
The optimality conditions of (\ref{ADMM:x}) and (\ref{ADMM:u}) are
\begin{subequations}
    \begin{align}
        \nabla_{\xx}L_\rho (\xx^{k+1}, \vv^k, \lamb^k) & = \PP^T \left[\lamb^k + \rho \left(\PP \xx^{k+1} - \vv^k\right)\right] = 0, \label{optim:x}\\
        \nabla_{\uu}L_\rho (\xx^{k+1}, \uu^{k+1}, \lamb^k) &= \nabla f(\uu^{k+1}) - \lamb^k - \rho \left(\PP \xx^{k+1} - \uu^{k+1} \right) = 0. \label{optim:u}
    \end{align}
\end{subequations}
According to optimality conditions (\ref{optim:x}), (\ref{ADMM:x}) can be expressed explicitly as (\ref{x}). 
\begin{subequations} \label{ADMM}
    \begin{align}
        \xx^{k+1} & = \PP^T\left(\vv^{k} - \frac{\lamb^k}{\rho}\right), \label{x}\\
        \uu^{k+1} & = \FF^{-1}\left( \frac{\sqrt{\bb^2+\varepsilon \mathbf{1}}}{\sqrt{\left| \FF\left(\PP\xx^{k+1} + \frac{\lamb^k}{\rho} \right) \right|^2+\varepsilon \mathbf{1}}}\odot \FF\left(\PP\xx^{k+1} + \frac{\lamb^k}{\rho}\right) \right). \label{u}
    \end{align}
\end{subequations}
However, it is difficult to obtain an explicit solution of the implicit equation (\ref{optim:u}). Thus we first use
$\tilde{\uu}^k$ to approximate $\uu^{k+1}$, then the gradient descent algorithm is adopted to solve the expression of $\uu^{k+1}$. According to (\ref{x}), we have
\begin{equation} \label{til_u}
    \vv^k = \PP \xx^{k+1} + \frac{\lamb^k}{\rho},
\end{equation}
where we set $\vv^k_{[n+1:m]} = \lamb^k_{[n+1:m]}/\rho$.
Then it's straightforward to show that $\nabla_{\uu}L_\rho (\xx^{k+1}, \vv^k, \lamb^k) = \nabla f(\vv^k) - \lamb^k - \rho \left(\PP \xx^{k+1} - \vv^k \right) = \nabla f(\vv^k)$. By gradient descent, it follows that
\begin{equation} \label{u_k+1}
    \uu^{k+1} = \vv^k - \nabla_{\uu}L_\rho (\xx^{k+1}, \vv^k, \lamb^k) = \vv^k - \nabla f(\vv^k).
\end{equation}
Substituting (\ref{equ5}) and (\ref{til_u}) into the expression of $\uu^{k+1}$ gives rise to (\ref{u}). 

Note that there is already algorithm \cite{Wen2012} for solving FPR problem using ADMM, we can also only solve sub-problem (\ref{sub1}) with ADMM, but adding an untrained generative network can reduce the requirement for sampling rates. Since the untrained generative network has a self-regularizing property \cite{Heckel2020}, i.e. the network recovers a natural image
at low sampling rate when trained with gradient descent until convergence. Therefore, We use the untrained generative network to project the estimated signal $\PP^T\hat{\uu}$ of ADMM to the range $\mathbb{S}$, getting the Net-ADM algorithm.

\begin{algorithm}[htbp]
    \renewcommand{\algorithmicrequire}{\textbf{Input:}}
	\renewcommand\algorithmicensure {\textbf{Output:} }
    \caption{Net-ADM}
    \begin{algorithmic}
        \REQUIRE $\bb, \vv^0 = \FF^{-1} ( \bb ), \lamb^0 = \mathbf{0}, \rho, \varepsilon$, K\\
        \ENSURE $\xx^K$ \\
    \end{algorithmic}
    \vskip 2mm
    \hrule
    \vskip 2mm
    \begin{algorithmic}[1]
        \STATE $\mathbf{for}\quad k = 0$ : K (outer loop) \\
        \STATE \quad $\xx^{k+1} = \PP^T \left(\vv^{k} - \frac{\lamb^k}{\rho} \right)$ \hfill (\ref{x})\\ 
        \STATE \quad $\uu^{k+1} = \FF^{-1}\left( \frac{\sqrt{\bb^2 + \varepsilon \mathbf{1}}}{\sqrt{|\FF (\PP \xx^{k+1} + \frac{\lamb^k}{\rho})|^2 + \varepsilon \mathbf{1}}} \odot \FF(\PP \xx^{k+1} + \frac{\lamb^k}{\rho}) \right) $ \hfill (\ref{u})\\
        \STATE \quad $\lamb^{k+1} = \lamb^k + \rho \left(\PP \xx^{k+1} - \uu^{k+1}\right)$ \hfill (\ref{ADMM:lambda})\\
        \STATE \quad $\ww^{k+1} = \underset{\ww}{\mathop{\arg\min}} \Vert \PP^T\uu^{k+1} - \GG(\ww;\zz) \Vert$  (inner loop) \hfill (\ref{ADMM:weight})\\
        \STATE \quad $\PP^T \vv^{k+1} = \GG(\ww^{k+1};\zz)$ \hfill (\ref{ADMM:net})\\
        \STATE $\mathbf{end}$
    \end{algorithmic}
\end{algorithm}

Since the model (\ref{equ13}) is non-convex, it is hard to ensure the algorithm converges to the optimal solution $\xx^{\ast} = \PP^T \uu^{\ast}$. Although the ADMM algorithm has a theoretical guarantee \cite{Lou2018} of convergence to a stable point under some conditions, the stable point may be far from the optimal solution due to the non-convexity of the model. We will show that the two projections included in the Net-ADM algorithm have good properties, one of which makes the objective function $f(\uu)$ descent and the other gets the estimation closer to the optimal solution.

\begin{figure}[H]
    \centering
    \includegraphics[height = 5cm]{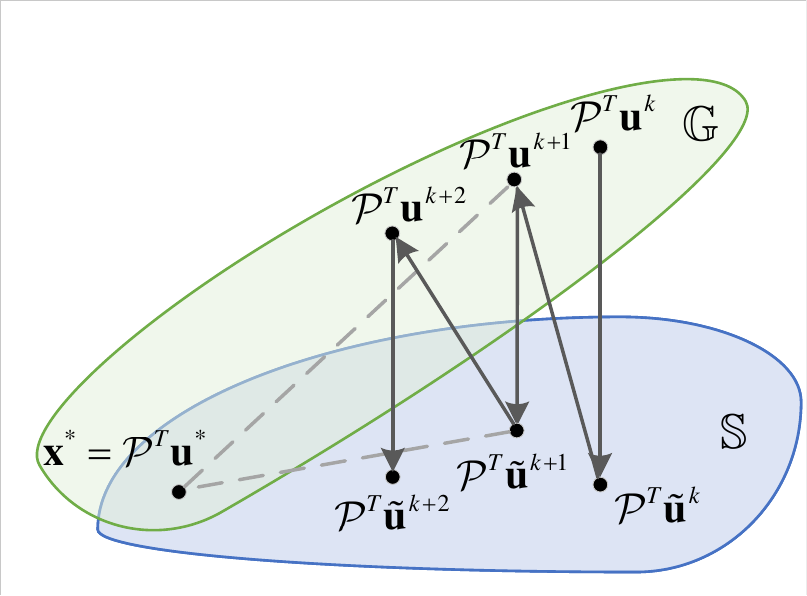}
    \caption{Image space visualization of FPR with untrained generative prior. The Net-ADM algorithm includes alternate projections of the sequences $(\PP^T\uu^k, \PP^T\vv^k)$ in two spaces. Where $\mathbb{S}=\{\GG(\ww;\zz):\ww \in \mathbb{W}, \GG(\ww;\zz) \in \RR^n\}$ denotes the generative space, which elements are obtained by the projections (\ref{ADMM:weight})-(\ref{ADMM:net}). $\mathbb{G} = \left\{\PP^T g(\vv):\vv \in \RR^m \right\}$ represents the set which elements are projected through the function $\PP^Tg:\RR^m \to \RR^n$. Note that $\xx^{\ast}=\PP^T\uu^{\ast} \in \mathbb{S} \cap \mathbb{G}$ since $\xx^{\ast} = \PP^T\uu^{\ast} = \PP^Tg(\uu^{\ast})$ and $\xx^{\ast} \in \mathbb{S}$. Additionally, we can also get $\PP^T\vv^{k+1} - \PP^T\uu^{k+1}$ is orthogonal to $\mathbb{S}$.}
    \label{fig: projection}
\end{figure}

Since the generative space $\mathbb{S}=\{\GG(\ww;\zz):\ww \in \mathbb{W}\}$ captures some characteristics of the images, we assume that the optimal solution is in the generative space $\xx^{\ast} \in \mathbb{S}$. Here we study the mutual projections of the sequences $(\PP^T \uu^k, \PP^T \vv^k)$. Let $g(\vv^k) = \vv^k - \nabla f(\vv^k)$, then
\begin{equation} \label{ite_u}
    \uu^{k+1} = g(\vv^k).
\end{equation}
If we set $\mathbb{G} = \left\{ \PP^T g(\vv):\vv \in \RR^m \right\}$, then the points in $\mathbb{S}$ can be projected into $\mathbb{G}$ through setting $\vv^k_{[n+1:m]} = \lamb^k_{[n+1:m]}/\rho$ and the function $\PP^T g:\RR^m \to \RR^n$. And (\ref{ADMM:weight}) and (\ref{ADMM:net}) are projections from $\mathbb{G}$ to $\mathbb{S}$. Thus, given $\PP^T\vv^0$ (note that although $\PP^T\vv^0$ may not be in $\mathbb{S}$, $\PP^T\uu^1$ is in $\mathbb{G}$), the mappings between sequences $\PP^T\uu^k$ and $\PP^T\vv^k \mbox{ } (k = 1,2,\cdots, K)$ in Net-ADM can be viewed as the alternate projections between $\mathbb{S}$ and $\mathbb{G}$, as shown in Fig.\ref{fig: projection}. Note that $\xx^{\ast} \in \mathbb{S} \cap \mathbb{G}$ since $\xx^{\ast} = \PP^T \uu^{\ast} = \PP^T g(\uu^{\ast})$ and $\xx^{\ast} \in \mathbb{S}$.

Here we can get some good properties about the projections.

\begin{theorem}
    Assume the optimal solution of (\ref{equ13}) is in the generative space $\xx^{\ast} \in \mathbb{S}$, the following statements are satisfied:\\
    (i) $f(\uu^{k+1}) \leq f(\vv^k), k = 0, 1, \cdots, K$, the equal sign holds only when $\nabla f(\vv^k)=0$.\\
    (ii) $\Vert \PP^T \vv^{k+1} - \xx^{\ast} \Vert \le \Vert \PP^T \uu^{k+1} - \xx^{\ast} \Vert, k = 0, 1, \cdots, K$, the equal sign holds when $\PP^T \uu^{k+1}$ happens to be in $\mathbb{S}$.\\
\end{theorem}
\begin{pof}
    (i) Note that projection (\ref{u}) is equivalent to (\ref{u_k+1}), which is the gradient descent on $f(\uu)$, it indicates that $f(\uu^{k+1}) \leq f(\vv^k)$, the equal sign holds only when $\nabla f(\vv^k)=0$. Moreover, it's obvious that $\uu^{\ast} = g(\uu^{\ast})$, thus according to Lagrange's mean value theorem we get
    \begin{equation}
        \uu^{k+1} - \uu^{\ast} = g(\vv^k) - g(\uu^{\ast}) = \langle \nabla g(\vv_\xi^k), \vv^k - \uu^{\ast} \rangle,
    \end{equation}
    where $\vv_\xi^k$ is a point between $\vv^k$ and $\uu^{\ast}$. Thus it's easy to get the following inequality
    \begin{equation}
        \Vert \uu^{k+1} - \uu^{\ast} \Vert \le \Vert \nabla g(\vv_\xi^k) \Vert \Vert \vv^k - \uu^{\ast} \Vert,
    \end{equation}
    if $\Vert \nabla g(\vv_\xi^k) \Vert < 1$, then $\Vert \uu^{k+1} - \uu^{\ast} \Vert < \Vert \vv^k - \uu^{\ast} \Vert$.
    
    (ii) The projections (\ref{ADMM:weight}) and (\ref{ADMM:net}) make
    \begin{equation}
        \Vert \PP^T \vv^{k+1} - \xx^{\ast} \Vert \le \Vert \PP^T \uu^{k+1} - \xx^{\ast} \Vert,
    \end{equation}
    because $\PP^T \vv^{k+1} - \PP^T \uu^{k+1}$ is orthogonal to $\mathbb{S}$. If $\PP^T \uu^{k+1}$ happens to be in $\mathbb{S}$, i.e. $\PP^T \uu^{k+1} \in \mathbb{S}$, then $\Vert \PP^T \vv^{k+1} - \xx^{\ast} \Vert = \Vert \PP^T \uu^{k+1} - \xx^{\ast} \Vert$. Else, $\PP^T \uu^{k+1} \notin \mathbb{S}$, then $\Vert \PP^T \vv^{k+1} - \xx^{\ast} \Vert < \Vert \PP^T \uu^{k+1} - \xx^{\ast} \Vert$.
\end{pof}

\begin{remark}
    In Net-ADM, the projection (\ref{u}) makes the objective function $f(\uu)$ descent, and projections (\ref{ADMM:weight}) and (\ref{ADMM:net}) make the estimation closer to the optimal solution $\xx^{\ast}$.
    Due to the non-convexity of model (\ref{equ13}), $f(\uu)$ may have multiple stable points, even more at low sampling rates. If we only use ADMM algorithm, it can indeed converge to a stable point under certain conditions. Although it may achieve superior results when the sampling rates are high, but it may become worse rapidly as the sampling rates decrease. Introducing the untrained generative prior can make the point closer to the optimal solution after reducing the objective function (or reaching a local minimum point), thereby reducing the probability of falling into a local minimum point.
\end{remark}

\section{Experiments} \label{experiment}
In this section, we compare the performance of Net-ADM with its closely related algorithms including ADMM, Net-GD and Net-PGD \cite{Jagatap2019, jagatapGauri2019} by numerical experiments. Note that there are a few experiments \cite{Wen2012} showing that ADMM performs better than some classical algorithms and it is competitive with the nonlinear conjugate gradient for related phase retrieval problems, so we only compare with ADMM in classic algorithms here. We mainly designed two groups of experiments, one is a comparison of the four algorithms under different sampling rates; the other is under different noise levels. Finally, we discuss the effects of parameters in Net-ADM.

\textbf{Dataset setup:} In order to fully compare the performance of the algorithms, we use images of different sizes and dimensions in the experiments. We select 6 grayscale images of $28\times 28$ from MNIST, and 5 RGB images center-cropped to $64\times 64\times 3$ from CelebA. In order to avoid the possible flip ambiguity of MNIST, we make a pre-process which sets the value of the two pixels near the edge of the image to 255, and their original values are 0. We also use a grayscale Cameraman image center-cropped to $128 \times 128$ as an example of large-size image.

\textbf{Network architecture:} We use the same network architecture as Net-GD and Net-PGD, 2-layer $\{25, 15, 10\}$ network is designed for MNIST and a 3-layer $\{120, 25, 15, 10\}$ network for CelebA. In addition, a 3-layer $\{128, 64, 64, 32\}$ network is designed for Cameraman.

\textbf{Measurement setup:} (i) We recover the unknown signals from Fourier magnitude measurements at different sampling rates $r = m/n$ which are the ratios of the measurement length to image length in each dimension. Note that the RGB image are considered as a stack of three two-dimensional images. (ii) With fixed sampling rates, we recover the unknown signals under the Fourier magnitude measurements affected by Gaussian noise of different levels.

\textbf{Performance metrics:} We compare the Peak Signal to Noise Ratio (PSNR) and Structural Similarity (SSIM) of the images reconstructed by ADMM, Net-GD, Net-PGD and Net-ADM under different sampling rates and under different noise levels respectively.

\textbf{Implementation details:} All algorithms use the Pytorch framework with Python 3. ADMM does not require any generative network, and the number of iterations is set to 5000. Net-GD, Net-PGD and Net-ADM have untrained generative networks, all networks in the algorithms use the Adam optimizer. The number of iterations of Net-GD is set to 5000, the learning rate is set to 0.005, which decays once every 2500 steps. Net-PGD and Net-ADM use network as the inner loop, we set the number of outer loops to $K = 1000$, the number of inner loops to 5. For Net-PGD, we set the learning rates of the outer loops and the inner loops to 0.5 and 0.0005 respectively, which decay once every 500 steps. For Net-ADM, we set the learning rate of outer loops to 0.005, which decays once every 500 steps. Regarding the setting of parameters in ADMM and Net-ADM, we set $\rho$ to 1 and $\varepsilon$ to 0.001 in all the experiments.

\subsection{FPR at different sampling rates}

To compare the reconstruction performance of ADMM, Net-GD, Net-PGD and Net-ADM algorithms at different sampling rates, we implement FPR at ten sampling rates (ranging from $1.1:0.1:2.0$) on the MNIST and CelebA respectively. Each algorithm was run ten times for each image at each sampling rate, and the average PSNR and SSIM of the reconstructed images were calculated as the mean of these ten runs.

\begin{figure}[htbp]
    \centering  
    \subfigure[PSNR of reconstructed images on MNIST]{
    \label{Fig1.sub.1}
    \includegraphics[width=0.48\textwidth]{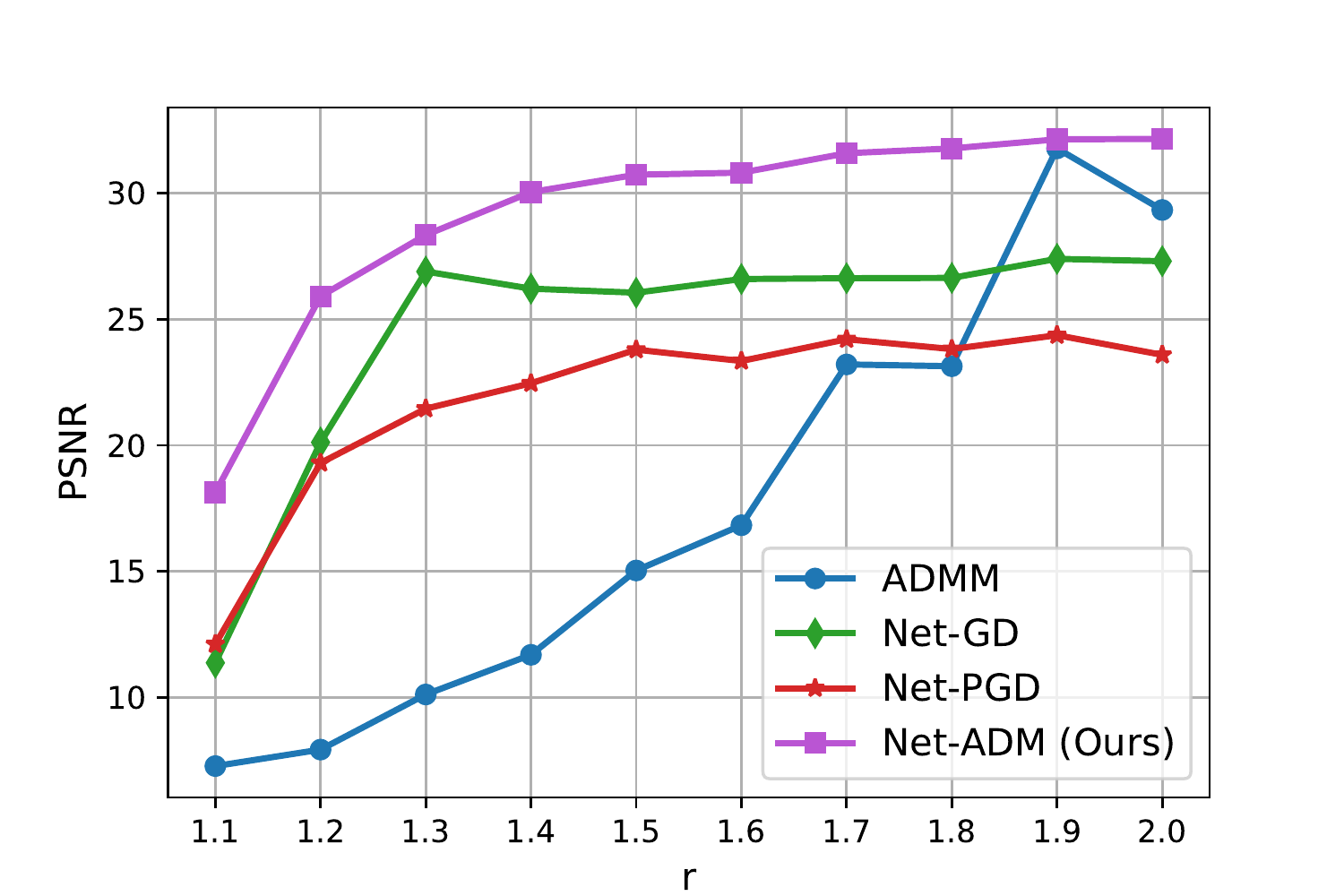}}
    \subfigure[SSIM of reconstructed images on MNIST]{
    \label{Fig1.sub.2}
    \includegraphics[width=0.48\textwidth]{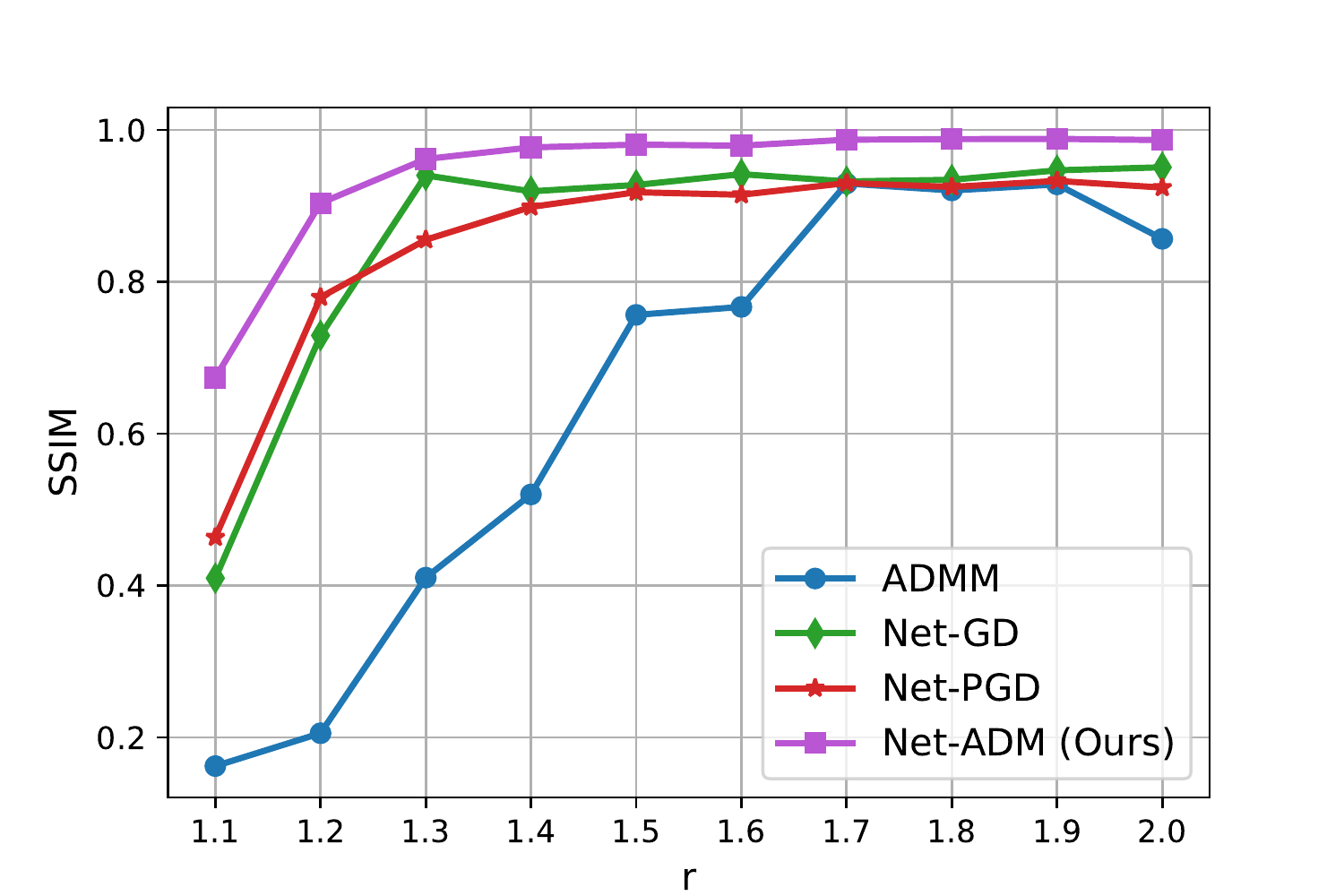}}
    \subfigure[PSNR of reconstructed images on CelebA]{
    \label{Fig1.sub.3}
    \includegraphics[width=0.48\textwidth]{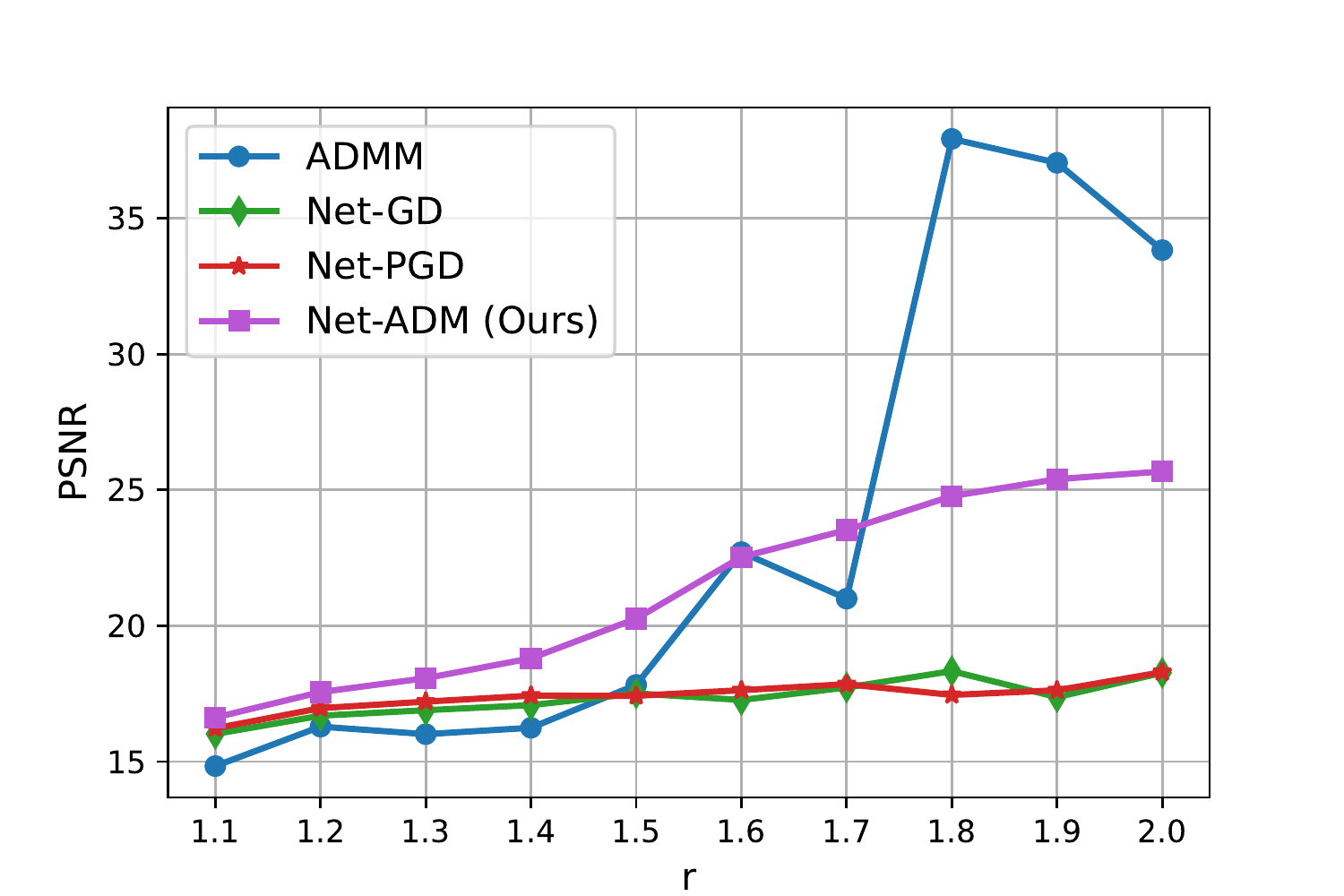}}
    \subfigure[SSIM of reconstructed images on CelebA]{
    \label{Fig1.sub.4}
    \includegraphics[width=0.48\textwidth]{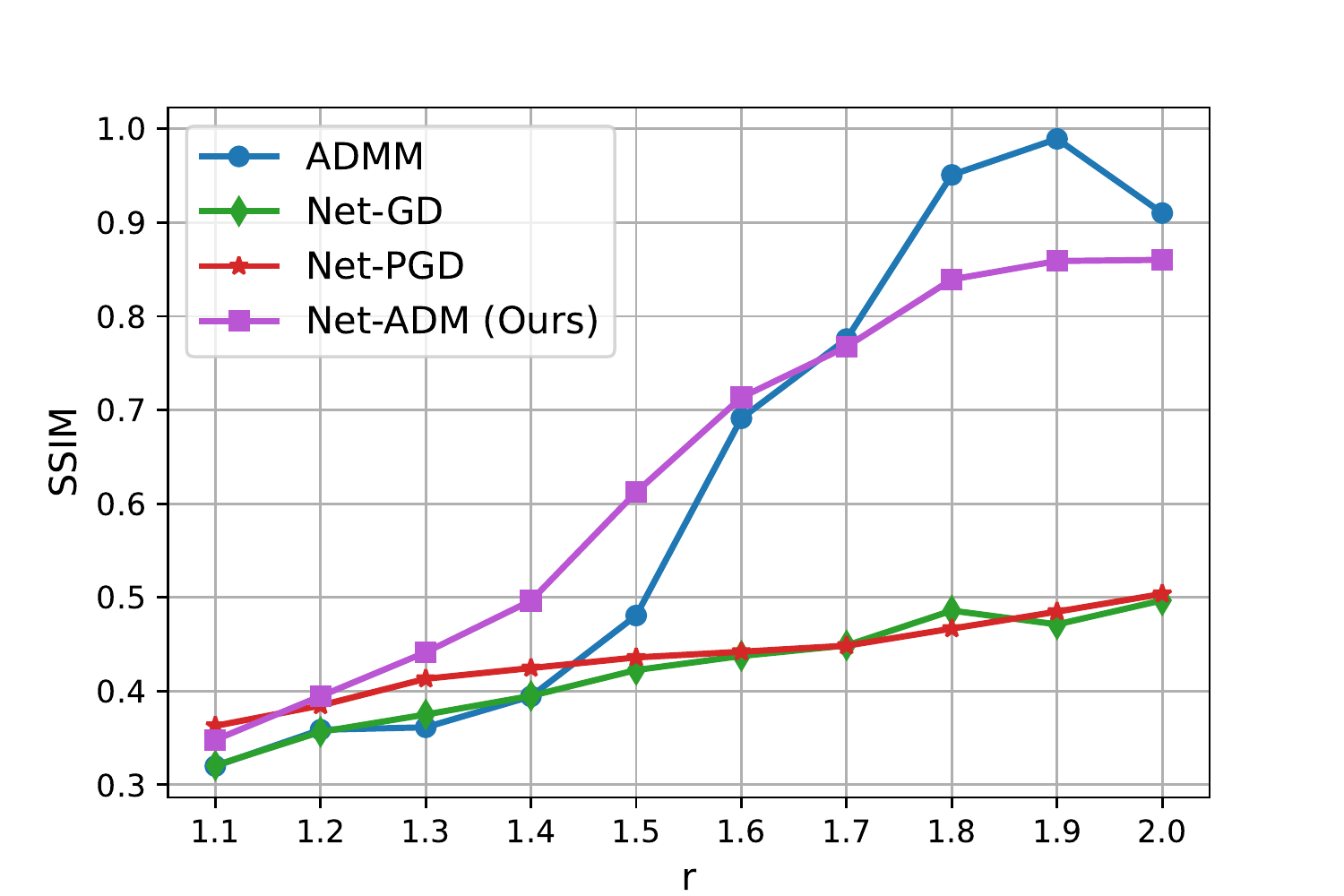}}
    \caption{The average PSNR and SSIM of reconstructed images on MNIST ($28\times 28$) and CelebA ($64\times 64 \times 3$). Each algorithm is run ten times for each image at each sampling rate ($r$), which are the ratios of measurement length to the image length in each dimension.}
    \label{fig:fig1}
\end{figure}

\begin{figure}[htbp]
    \centering
    \subfigure[Reconstructed images on MNIST at sampling rates $r = 1.1:0.1:1.6$]{
    \label{Fig2.sub.1}
    \includegraphics[height=5.5cm]{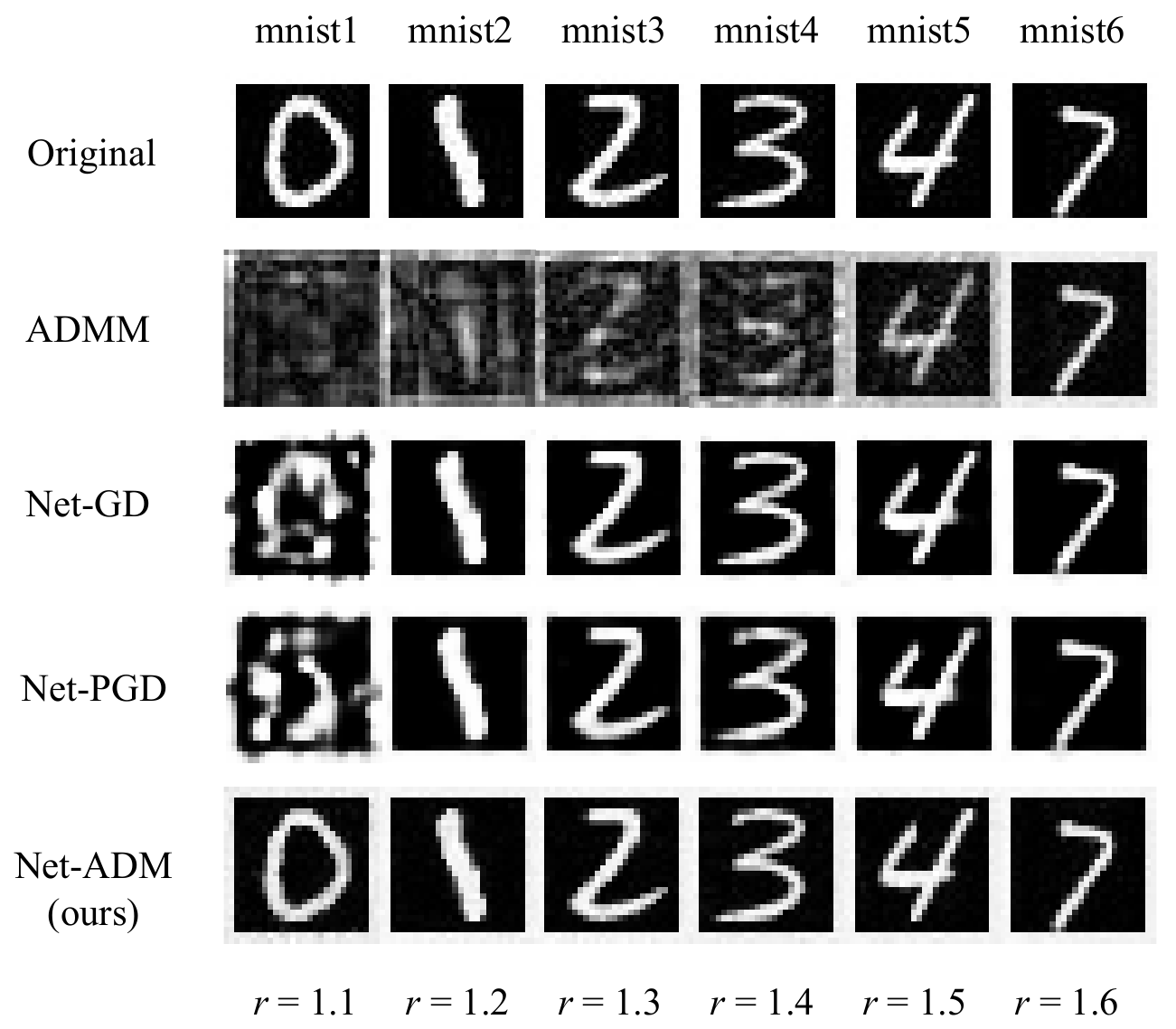}}
    \subfigure[Reconstructed images on CelebA at sampling rates $r = 1.6:0.1:2.0$]{
    \label{Fig2.sub.2}
    \includegraphics[height=5.5cm]{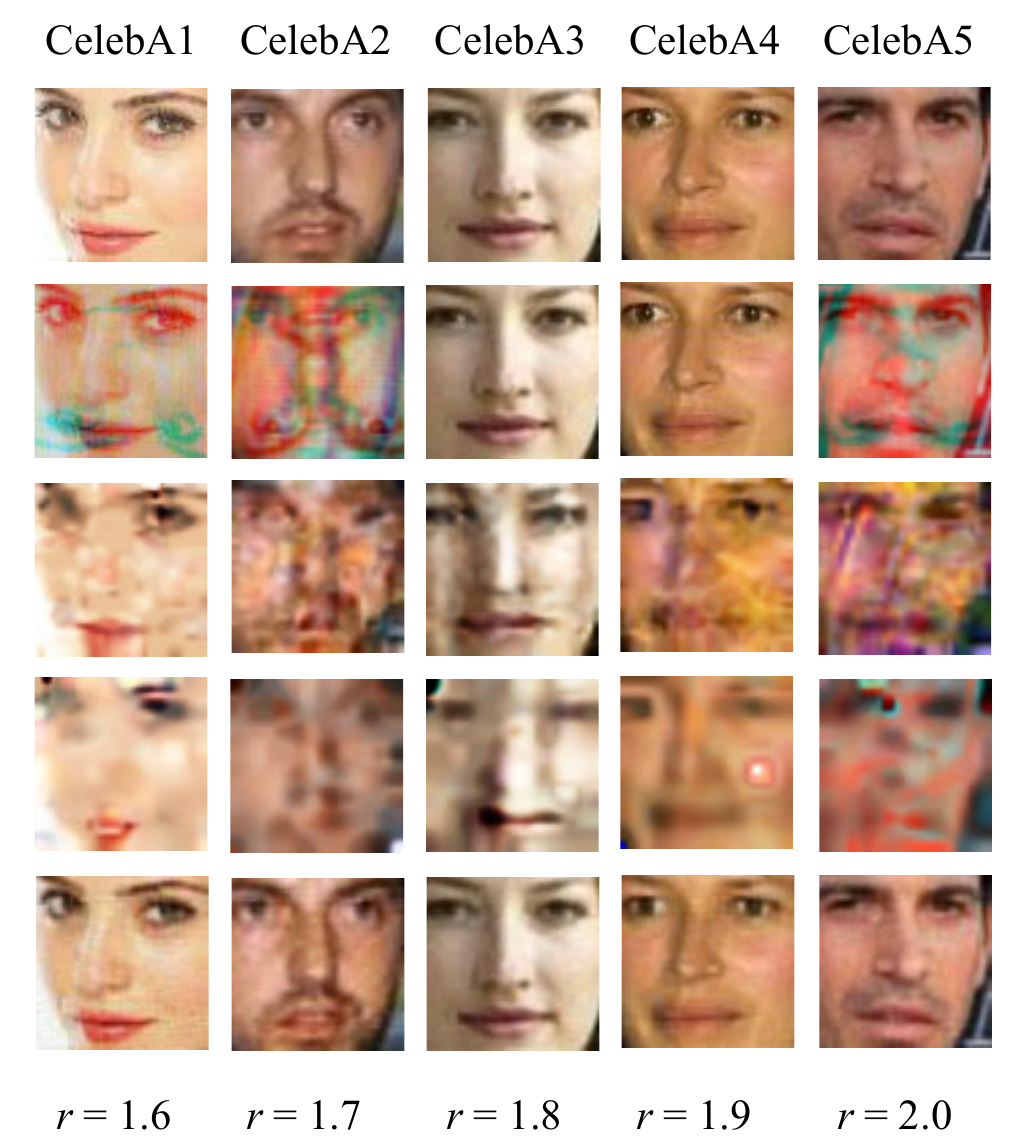}}
    \caption{The reconstructed results on MNIST ($28 \times 28$) and CelebA ($64\times 64 \times 3$). Each algorithm is run ten times for each image at each sampling rate ($r$), and the best results in ten runs are shown here. The images in MNIST and CelebA are reconstructed at sampling rates $r = 1.1:0.1:1.6$ and $r = 1.6:0.1:2.0$ respectively.}
    \label{fig:fig2}
\end{figure}

Fig.\ref{fig:fig1} shows the average PSNR and SSIM of the images reconstructed by the four algorithms at different sampling rates. The results show that the average PSNR and SSIM of Net-ADM's reconstructed results are higher than those of the other three algorithms in most instances, especially when $r < 1.7$. Although the average PSNR and SSIM of ADMM's reconstructed results might be higher than those of Net-ADM when $1.7 < r \le 2.0$, they decreases rapidly when the sampling rates are lower than 1.7 (for MNIST) or 1.8 (for CelebA), which shows that ADMM has higher requirements for sampling rates. When $r > 1.7$, the reason why Net-ADM is sometimes slightly worse than ADMM may be that the number of parameters of the untrained generative network is not enough to accurately fit the high-quality estimation $\PP^T\uu$ generated by ADMM. After all, we have to make a trade-off between fitting error and stability. Comparatively, Net-GD, Net-PGD and Net-ADM are more stable with the decrease of sampling rates, which is in fact more challenging for FPR especially in applications where the resolution of the measurement device is low. Among them, the average PSNR and SSIM of images reconstructed by Net-ADM are higher than those of Net-GD and Net-PGD.

\begin{figure}
    \centering
    \includegraphics[height = 7.5cm]{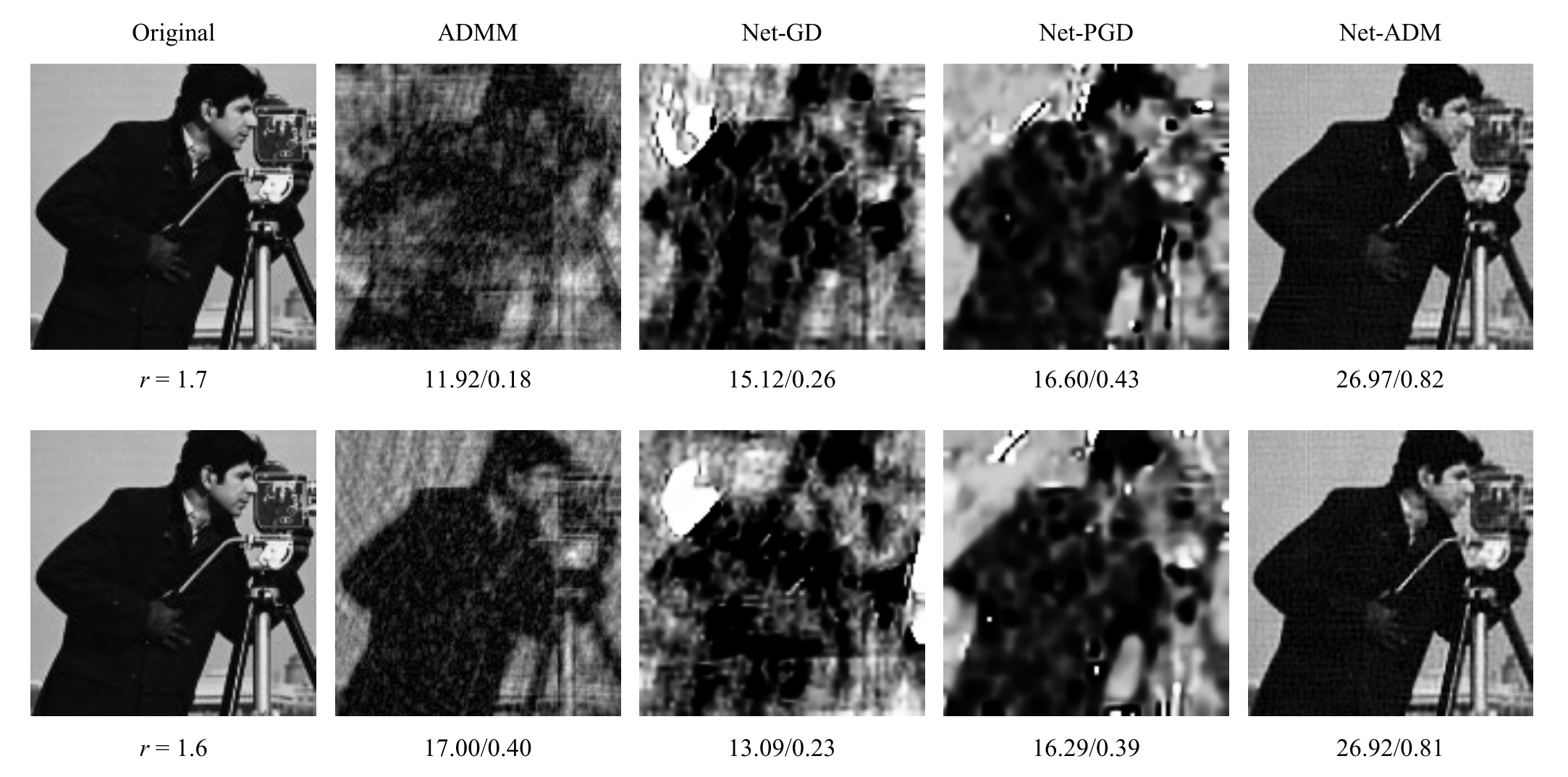}
    \caption{The reconstructed results on Cameraman ($128 \times 128$). Each algorithm is run ten times for each image at each sampling rate ($r$), and the best results in ten runs are shown here. The images in the first row are reconstructed results at sampling rate $r = 1.7$, and the second row are results at sampling rate $r = 1.6$. The PSNR and SSIM of reconstructed results are indicated below the images, they are denoted by PSNR(dB)/SSIM.}
    \label{fig:fig3}
\end{figure}

Fig.\ref{fig:fig2} shows the best reconstructed images (on MNIST and CelebA) in ten runs at sampling rates $r = 1.1:0.1:1.6$ and $r = 1.6:0.1:2.0$ respectively. Fig.\ref{fig:fig3} shows the best reconstructed images on Cameraman from ten runs of the four algorithms at sampling rate $r = 1.7$ and $r = 1.6$ respectively. From the experiment results, the reconstruction quality using Net-ADM is superior to others. Although ADMM seems to work better when the sampling rate is high, it's unstable such as the reconstruction quality of CelebA5 at $r = 2.0$ is poor. Additionally, the reconstructed results of Net-GD and Net-PGD are not as good as Net-ADM. Thus, the reconstructed results of Net-ADM at different sampling rates are superior and stable by contrast.

\subsection{FPR at different noise levels}

In this section, we compare the robustness of Net-ADM with other algorithms in resisting measurement noise. The noisy measurement model is:
\begin{equation}
    \bb = \left| \FF\xx \right| + \eta
\end{equation}
where $\eta$ is Gaussian noise, i.e. $\eta \sim N(0,\sigma^2)$.

Under fixed sampling rates, we compare the PSNR and SSIM of the images reconstructed by four algorithms when the measurements are affected by noise of different levels. The noise level is measured by Signal Noise Ratio (SNR), the formula for calculating SNR is as follows:
\begin{equation}
    {\rm SNR} = 20 \log_{10}{\frac{{\rm Var}(\left| \FF\xx \right|)}{\sigma^2}}
\end{equation}
where ${\rm Var}(\left| \FF\xx \right|)$ calculates the variance of the noiseless magnitude measurement $\left| \FF\xx \right|$. At each SNR we will randomly generate 10 different Gaussian noise.

\begin{table}[htbp] \small
    \centering
    \caption{The average PSNR and SSIM of CelebA3's reconstructed results in ten runs when the magnitude measurements are effected by noise of different levels (SNR) under four fixed sampling rates ($r$).}
    \label{tab:tab1}
    \setlength{\tabcolsep}{1.2mm}{
    \begin{tabular}{c|c|c|c|c|c|c|c}
        \hline \hline
        \multicolumn{2}{c|}{SNR} & 70 & 60 & 50 & 40 & 30 & 20 \\
        \hline
        \multirow{4}{*}{$r$=2.0} & ADMM & 23.55/0.84 & 21.14/0.76 & 18.78/0.65 & 16.82/0.56 & 14.60/0.41 & 11.48/0.26 \\
        & Net-GD & 13.88/0.43 & 13.65/0.43 & 14.17/0.44 & 14.41/0.44 & 14.20/0.39 & 14.20/0.40  \\
        & Net-PGD & 17.25/0.54 & 14.74/0.43 & 16.30/0.51 & 15.88/0.52 & 15.16/0.47 & 14.23/0.43 \\
        & Net-ADM (Ours) & \textbf{27.79/0.90} & \textbf{26.12/0.87} & \textbf{25.03/0.86} & \textbf{21.47/0.75} & \textbf{19.83/0.64} & \textbf{17.36/0.50} \\
        \hline
        \multirow{4}{*}{$r$=1.8} & ADMM & 23.22/\textbf{0.85} & 20.81/0.74 &	19.03/\textbf{0.66} &	15.99/0.52 & 13.80/0.37 & 11.36/0.24 \\
        & Net-GD & 14.49/0.42 & 14.96/0.43 & 15.59/0.45 & 15.26/0.43 & 15.93/0.42 & 14.16/0.35 \\
        & Net-PGD & 17.84/0.56 & 15.34/0.43 & 15.61/0.42 & 14.82/0.48 & 14.06/0.39 & 14.51/0.39 \\
        & Net-ADM (Ours) & \textbf{24.86}/0.81 & \textbf{23.83/0.80} & \textbf{20.43/0.66} & \textbf{21.15/0.71} & \textbf{17.54/0.51} & \textbf{17.21/0.50} \\
        \hline
        \multirow{4}{*}{$r$=1.6} & ADMM & 22.46/\textbf{0.81} & 19.98/0.73 & 17.82/0.61 &	15.91/0.50 & 12.80/0.33 & 10.12/0.18 \\
        & Net-GD & 14.82/0.38 & 14.38/0.38 & 14.91/0.38 & 14.78/0.36 & 13.32/0.32 & 13.92/0.33 \\
        & Net-PGD & 14.64/0.41 & 14.84/0.42 & 13.88/0.43 & 16.61/0.48 & 15.69/0.42 & 13.66/0.34 \\
        & Net-ADM (Ours) & \textbf{24.67}/0.80 & \textbf{25.21/0.83} & \textbf{23.31/0.79} & \textbf{21.74/0.72} & \textbf{18.29/0.56} & \textbf{15.75/0.41} \\
        \hline
        \multirow{4}{*}{$r$=1.4} & ADMM & 12.64/0.33 & 12.52/0.32 & 12.34/0.31 & 11.86/0.28 & 11.41/0.23 & 9.53/0.16 \\
        & Net-GD & 14.55/0.36 & 14.68/0.36 & 12.93/0.32 & 14.02/0.36 & 13.77/0.33 & 13.16/0.28 \\
        & Net-PGD & 14.52/0.38 & 13.61/0.35 & 14.92/0.38 & 14.95/0.37 & 13.52/0.33 & 13.49/0.30 \\
        & Net-ADM (Ours) & \textbf{16.61/0.47} & \textbf{16.61/0.45} & \textbf{16.45/0.43} & \textbf{16.14/0.41} & \textbf{14.82/0.37} & \textbf{14.48/0.34} \\
        \hline \hline
    \end{tabular}}
\end{table}

\begin{figure}
    \centering
    \includegraphics[height = 7.5cm]{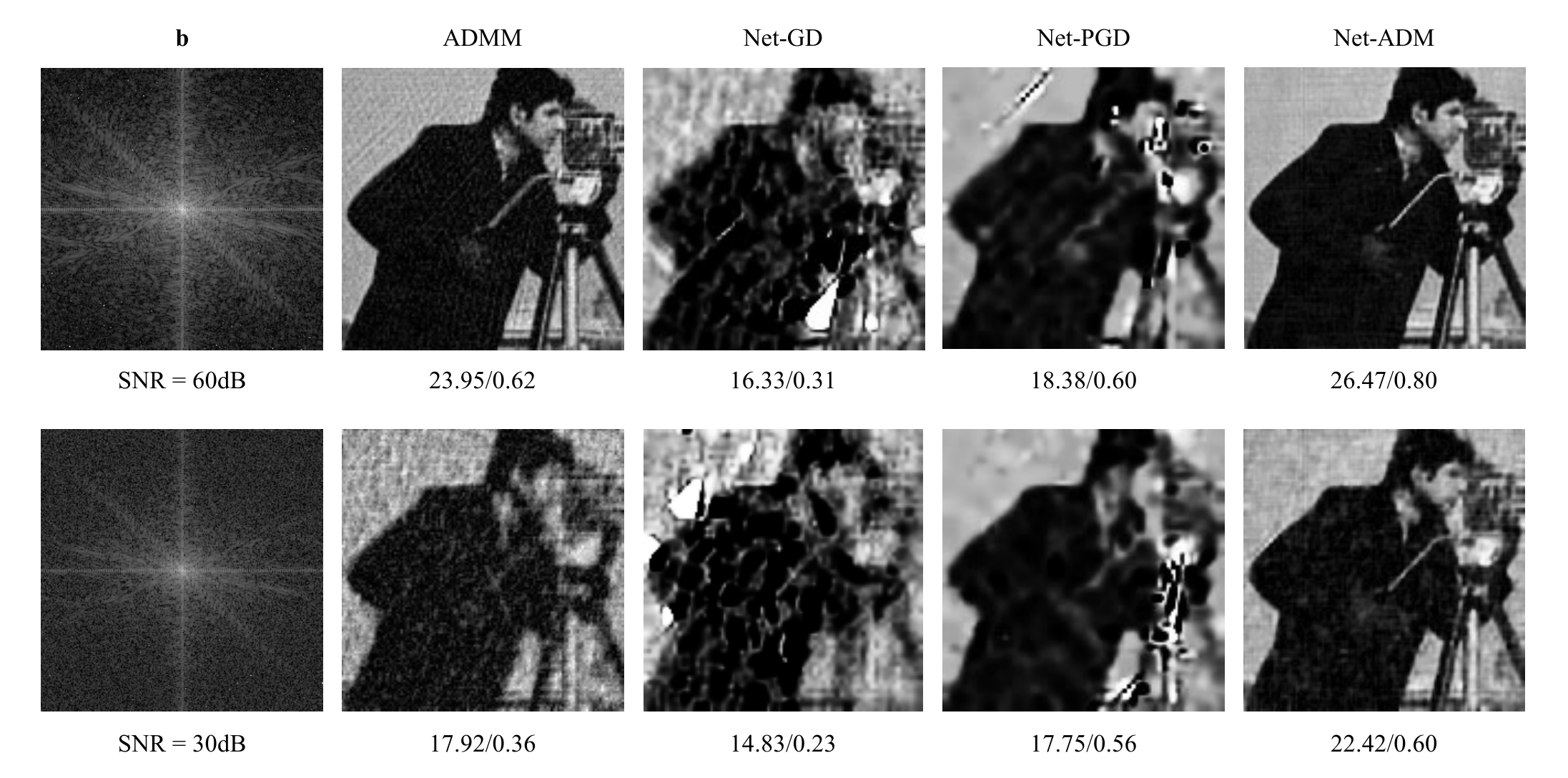}
    \caption{The reconstructed results on Cameraman ($128 \times 128$). Each algorithm is run ten times for each image at each SNR, and the best results in ten runs are shown here. The first row is the reconstructed images under SNR = 60dB, and second row is under SNR = 30dB. The first column shows the magnitude measurements affected by noise of different levels, when the sampling rate is 2.0. The PSNR and SSIM of reconstructed results are indicated below the images, they are denoted by PSNR(dB)/SSIM.}
    \label{fig:fig5}
\end{figure}

We choose CelebA3 image, add noise of different levels (SNR ranging from 20:10:70dB) to the measurements under four fixed sampling rates, and compare the PSNR and SSIM of the reconstructed results of the four algorithms. The average PSNR and SSIM of ten runs are shown in Table \ref{tab:tab1}.
In addition, we choose Cameraman image, add noise of SNR=60dB and SNR=30dB to the magnitude measurements respectively at sampling rate $r=2.0$, the best results in ten runs are shown in Fig.\ref{fig:fig5}.

It is observed from the experiment results that the PSNR and SSIM of Net-ADM's reconstructed results are basically higher than those of other three algorithms under different SNR and sampling rates. The robustness of Net-ADM is significantly improved compared to ADMM, even though the PSNR and SSIM of their reconstructed results decrease as the sampling rates and SNR decrease. The PSNR and SSIM of the images reconstructed by Net-GD and Net-PGD remains stable with the change of sampling rates and SNR, but Net-ADM's constructed results maintain the highest PSNR and SSIM in most situations, indicating that Net-ADM has superior robustness.

\subsection{Impact of parameters}

Now we consider the effect of parameters in Net-ADM. Because the ADMM is not highly sensitive to the choice of the relaxation parameter $\rho$ \cite{Wen2012}, we don't discuss too much about $\rho$ here.
Since the design of the architecture will affect the representation range of the network, we explore the important parameters that affect the network architecture. Due to the particular choice of the input $\zz$ is not very important \cite{Heckel2019} but it is necessary to ensure that the lines are not coherent, we won't explore exhaustively here. Upsampling is a vital part of network, it has been discussed in \cite{Heckel2019} that linear upsampling is better able to represent smoothly varying portions of the signal compared with no upsampling, nearest neighbor upsampling, convex and non-linear upsampling. Regarding other tunable parameters of the untrained generative network, we compare the reconstructed results of Net-ADM at different network depths, number of weight channels and activation functions. Net-ADM is run ten times at each sampling rate in all experiments.

\begin{figure}[H]
    \centering
    \subfigure[Impact of depth $J$]{
    \label{Fig6.sub.1}
    \includegraphics[width=0.48\textwidth]{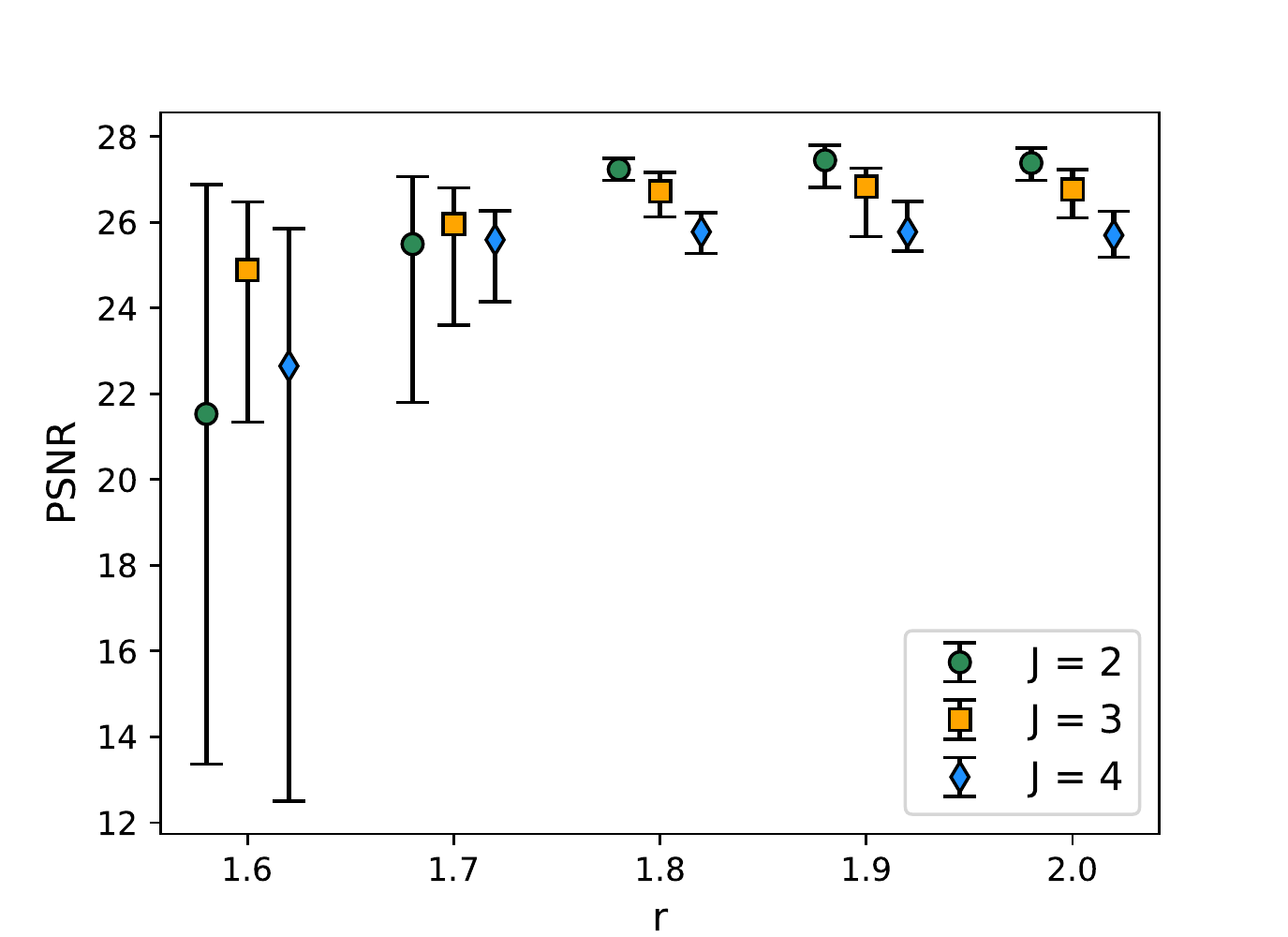}}
    \subfigure[Impact of channels \{$c_0, c_1, \cdots, c_J$\} ]{
    \label{Fig6.sub.2}
    \includegraphics[width=0.48\textwidth]{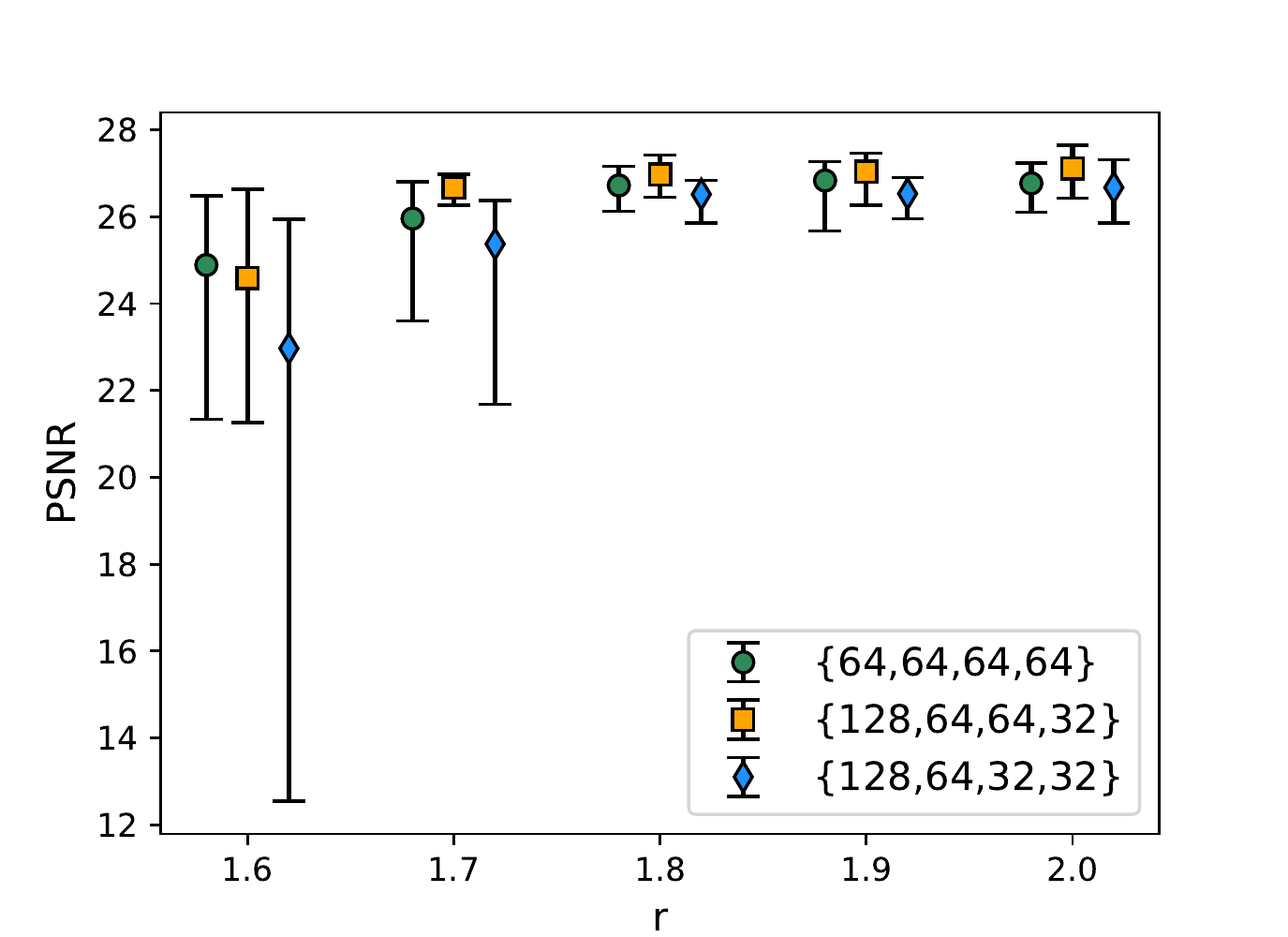}}
    \caption{(a): The PSNR of Net-ADM's reconstructed image (Cameraman) at different network depths when the number of channels per layer is 64. (b): The PSNR of Net-ADM's reconstructed image (Cameraman) at different numbers of channels when the network depth is 3. The figures show the average values in the PSNR interval out of ten runs.}
    \label{fig:fig6}
\end{figure}

In order to evaluate the impact of network depth, we fix the channel number in each layer of the network to 64 and the activation function to ReLU, and compare the reconstructed results of Net-ADM on Cameraman with different network depths. The PSNR of the reconstructed results are shown in Fig.\ref{fig:fig6} (a). Although the PSNR of reconstructed image under network depth $J = 3$ is low slightly than $J = 2$ sometimes, it remains stable interval in all situations. Taking the average quality and stability of the reconstruction into account, we choose the network depth to be 3. 

Then in the case where the network depth is 3 and the activation function is ReLU, we compare the impact of the number of weight channels, as shown in Fig.\ref{fig:fig6} (b). It can be seen that the PSNR of reconstructed image under different channel numbers is slightly different, but we choose the relatively optimal scheme, which is \{128, 64, 64, 32\}.

\begin{figure}[H]
    \centering
    \includegraphics[height=5.5cm]{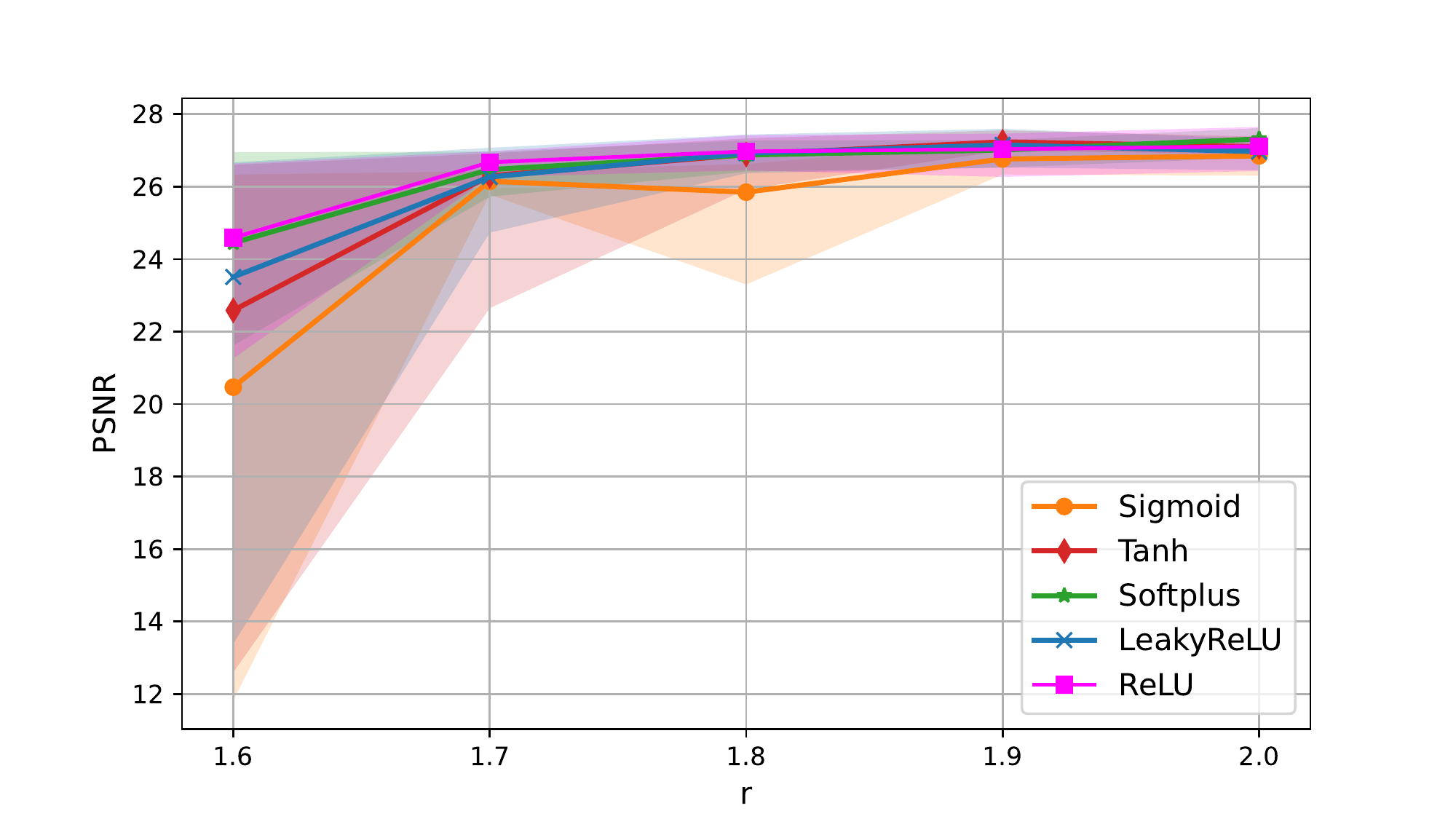}
    \caption{The PSNR of Net-ADM's reconstructed image (Cameraman) at different activation functions when the number of network channels are designed by $\{128, 64, 64, 32\}$. The figure shows the the average values in the PSNR interval out of ten runs.}
    \label{fig:fig7}
\end{figure}

In the case where the number of network channels is designed by \{128, 64, 64, 32\}, we reconstruct image under different activation functions, the average values in the PSNR interval out of ten runs are shown in Fig.\ref{fig:fig7}. Except for the Sigmoid activation function, the effects of these activation functions are not significantly different when sampling rates are high, but the opposite when sampling rates are low. Thus, we choose the ReLU activation function which is relatively stable.

\section{Conclusion} \label{conclusion}
In this paper, we propose an algorithm named Net-ADM which combines ADMM with untrained generative prior to solve FPR problem. We theoretically analyze that the projections included in the algorithm have good properties under certain conditions, one of which makes the objective function descent, and the other makes the estimation closer to the optimal solution. And we numerically prove the superiority of the algorithm, the reconstruction performance of Net-ADM is mostly superior to state-of-the-art algorithms at different sampling rates, especially at low sampling rates. In addition, the robustness to Gaussian noise is stronger than other algorithms. Finally, we discuss parameters that may affect the algorithm, network depth is an important factor and the number of channels and activation function are also relevant factors.

\bibliographystyle{elsarticle-num}
\bibliography{main}

\end{document}